\documentclass[twocolumn,superscriptaddress,aps,prl,amsmath,amssymb]{revtex4}
\usepackage{graphicx}
\usepackage{dcolumn}
\usepackage{bm}
\usepackage{epsfig}
\usepackage[usenames]{color}
\usepackage{amsthm} 
\usepackage[colorlinks=true,linkcolor=blue,citecolor=blue,urlcolor=blue]{hyperref}
\usepackage{setspace}
\usepackage{color}

\let\baraccent=\=

\renewcommand{\v}[1]{\ensuremath{\mathbf{#1}}}

\renewcommand{\=}[1]{\stackrel{#1}{=}}

\newcommand{\al}[1]{\begin{align}#1\end{align}}
\newcommand{\bs}{\begin{split}}
\newcommand{\es}{\end{split}}

\newcommand{\BiSb}{Bi$_{1-x}$Sb$_x$~}
\def\be{\begin{equation}}
\def\ee{\end{equation}}

\begin{document}
\title{Tunable Giant Spin Hall Conductivities in a Strong Spin-Orbit Semimetal: Bi$_{1-x}$Sb$_x$}
\author{C\"{u}neyt \c{S}ahin}
 \email{cuneyt-sahin@uiowa.edu}
\author{Michael E. Flatt\'e}
 \email{michael\_flatte@mailaps.org}
\affiliation{Optical Science and Technology Center and Department of Physics and Astronomy, University of Iowa, Iowa City, Iowa 52242, USA}
\date{\today}
\begin{abstract}

Intrinsic spin Hall conductivities are calculated for strong spin-orbit  Bi$_{1-x}$Sb$_x$ semimetals, from the Kubo formula and using Berry curvatures evaluated throughout the Brillouin zone from a 
tight-binding Hamiltonian. Nearly-crossing bands with strong spin-orbit interaction generate \hbox{giant} spin Hall conductivities in these materials, ranging from 474 ($\hbar$/e)($\Omega$cm)$^{-1}$ for bismuth to 96 ($\hbar$/e)($\Omega$cm)$^{-1}$ for antimony; the value for bismuth is more than twice that of platinum. The large spin Hall conductivities persist for alloy compositions corresponding to a three-dimensional topological insulator state, such as Bi$_{0.83}$Sb$_{0.17}$. The spin Hall conductivity could be changed by a factor of five for doped Bi, or for Bi$_{0.83}$Sb$_{0.17}$, by changing the chemical potential by $0.5$~eV, suggesting the potential for doping or voltage tuned spin Hall current.

\end{abstract}
\pacs{pacs numbers}

\maketitle
Spin currents flowing transverse to  electric fields, known as spin Hall currents, originate from spin-orbit interaction in a solid and the resulting spin-orbit entanglement of electronic states\cite{Engel2007,Murakami2005,Vignale2010}. The spin Hall conductivity, which is the ratio of the spin Hall current to the longitudinal  electric field,  depends on details of the electronic band structure such as the strength of the spin-orbit interaction, the Fermi energy, the direction of current relative to crystal axes and the strain\cite{Dyakonov1971,Hirsch1999,Murakami2003,Kato2004b,Sih2005,Guo2005,Yao2005,Hankiewicz2006,Sih2006,Guo2008,Lowitzer2011,Liu2012,Norman2014,Norman2014e}. Such dependencies may provide ways to electrically control the spin Hall conductivity. Measurements of the variation of the spin Hall conductivity with these quantities have been done in most detail for non-centrosymmetric semiconductor quantum wells\cite{Sih2005,Norman2014},  
however other phenomena, including current-induced spin polarization\cite{Kato2004a,Kato2005b} and precessional spin-orbit fields\cite{Kato2004c} (which also depend on the electronic band structure) complicate the interpretation. 
 Centrosymmetric metals have fewer complicating effects, and their so-called spin Hall angle (the ratio of the spin current to the longitudinal charge current) is much larger than for semiconductors\cite{Yao2005,Valenzuela2006,Vila2007,Guo2008,Liu2012} but it is considerably more difficult to modify the intrinsic conductivities of high conductivity metals. This suggests consideration of centrosymmetric semimetals with large spin-orbit couplings, such as bismuth and antimony, as these might have more tunable spin Hall conductivities and longitudinal conductivities while maintaining very large spin Hall angles. Large spin Hall angles have been demonstrated for bismuth selenide\cite{Mellnik2014}, motivated by proposals for large spin current effects in topological insulators\cite{Burkov2010,Culcer2010,Pesin2012}

Here we report calculations of  the intrinsic spin Hall effect for bismuth, antimony and bismuth-antimony alloys and find values more than two orders of magnitude larger than other voltage-tunable materials, and greatly in excess of all other measured materials except bismuth selenide. As the alloy composition changes, these materials change from semimetallic to three-dimensional topological insulating\cite{Hsieh2008,Hsieh2009} with a single Dirac cone, and back to semimetallic, but the bulk spin Hall conductivity varies smoothly through those transitions. We trace the origin of these effects to energy-resolved Berry curvature contributions to the spin Hall conductivity. The presence of nearly-overlapping bands with large spin-orbit interaction near the Fermi energy in these alloys produces a highly responsive dependence of the spin Hall conductivity on the Fermi energy or carrier density. We thus identify a class of materials in which giant spin Hall conductivities can be effectively tuned with modest voltages. These materials mirror ordinary semiconductors where the conductivity can be changed dramatically with a modest voltage; here the spin Hall conductivity demonstrates ``semiconducting behavior''.

Bismuth and antimony are both
semimetals 
with enormous spin-orbit couplings, 1.5 eV and 0.6 eV respectively\cite{Gonze1990}. 
These elements are both rhombohedral crystals with space group of $D_{3d}^5$ (R$\bar 3$m) and point group $D_{3d}$ ($\bar 3$m) \cite{Ast2003}. Their semimetallic behaviour comes from slightly overlapping conduction and valence bands resulting in electron pockets at the L points of the Brillouin zone and hole pockets at the T points for bismuth and the H points for antimony. The overlap between L and H is 180 meV in Sb and between L and T is 40 meV in Bi \cite{Issi1979}. A low-energy effective spin-orbit Hamiltonian with a third nearest-neighbor tight-binding parameterization\cite{Liu1995} suffices to mimic the characteristics of the electronic structure and the effective masses around the Fermi energy, electron, and hole pockets. For the electronic structure of the Bi$_{1-x}$Sb$_{x}$ alloy\cite{Bellaiche2000}, the band energies and overlap integrals are averaged using the virtual crystal approximation.  

\begin{figure}
\centering
 \includegraphics[width=.99\linewidth]{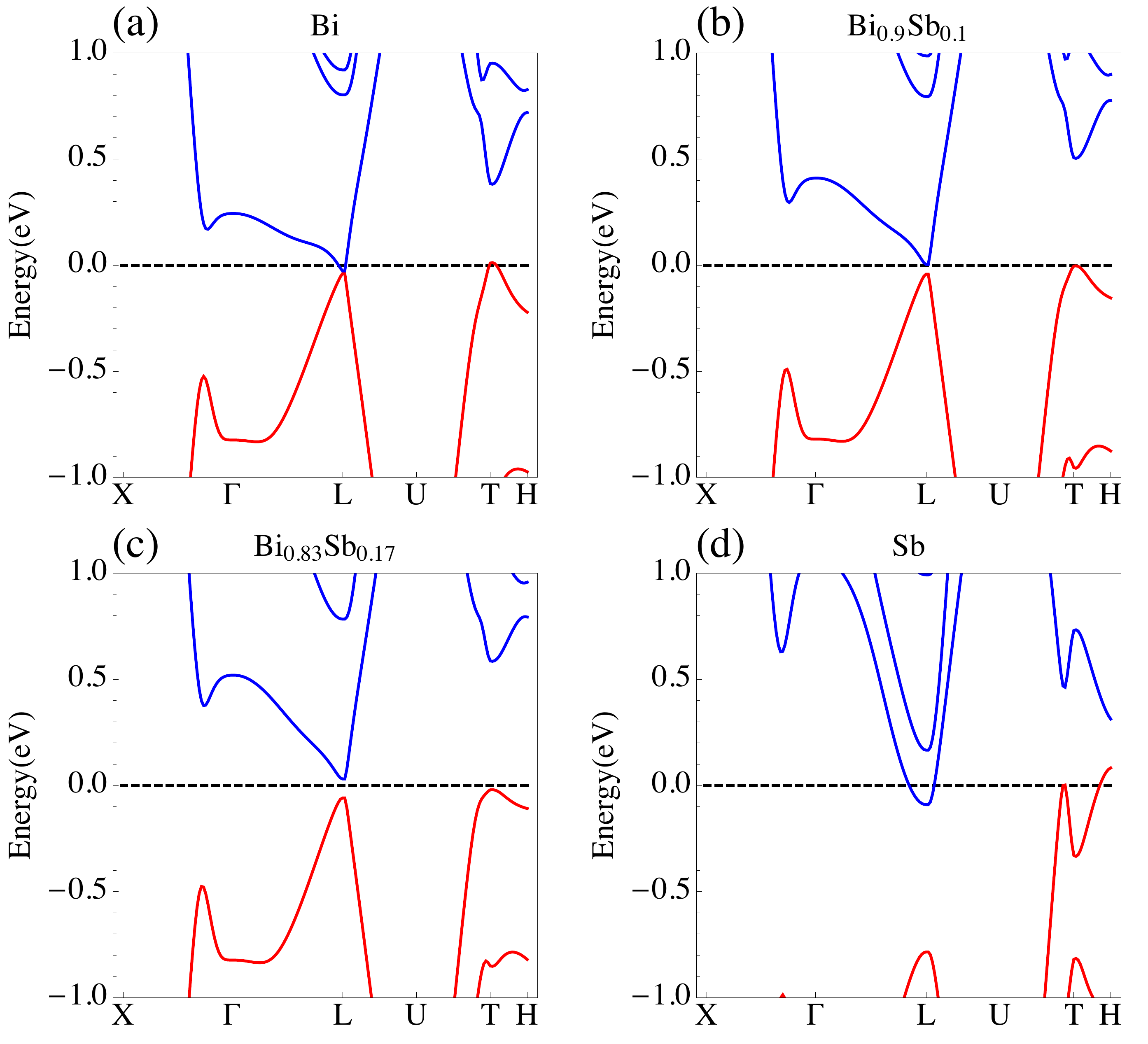} 
  \caption{Electronic band structure for (a) pure bismuth, (b) Bi$_{0.9}$Sb$_{0.1}$ with disappearing band overlap, (c) semiconducting Bi$_{0.83}$Sb$_{0.17}$ and (d) pure antimony. The Fermi level is at 0~eV for each.}
   \label{fig:bands}
\end{figure}

\begin{figure}
  \begin{minipage}[b]{0.49\linewidth}
    \includegraphics[width=.99\linewidth]{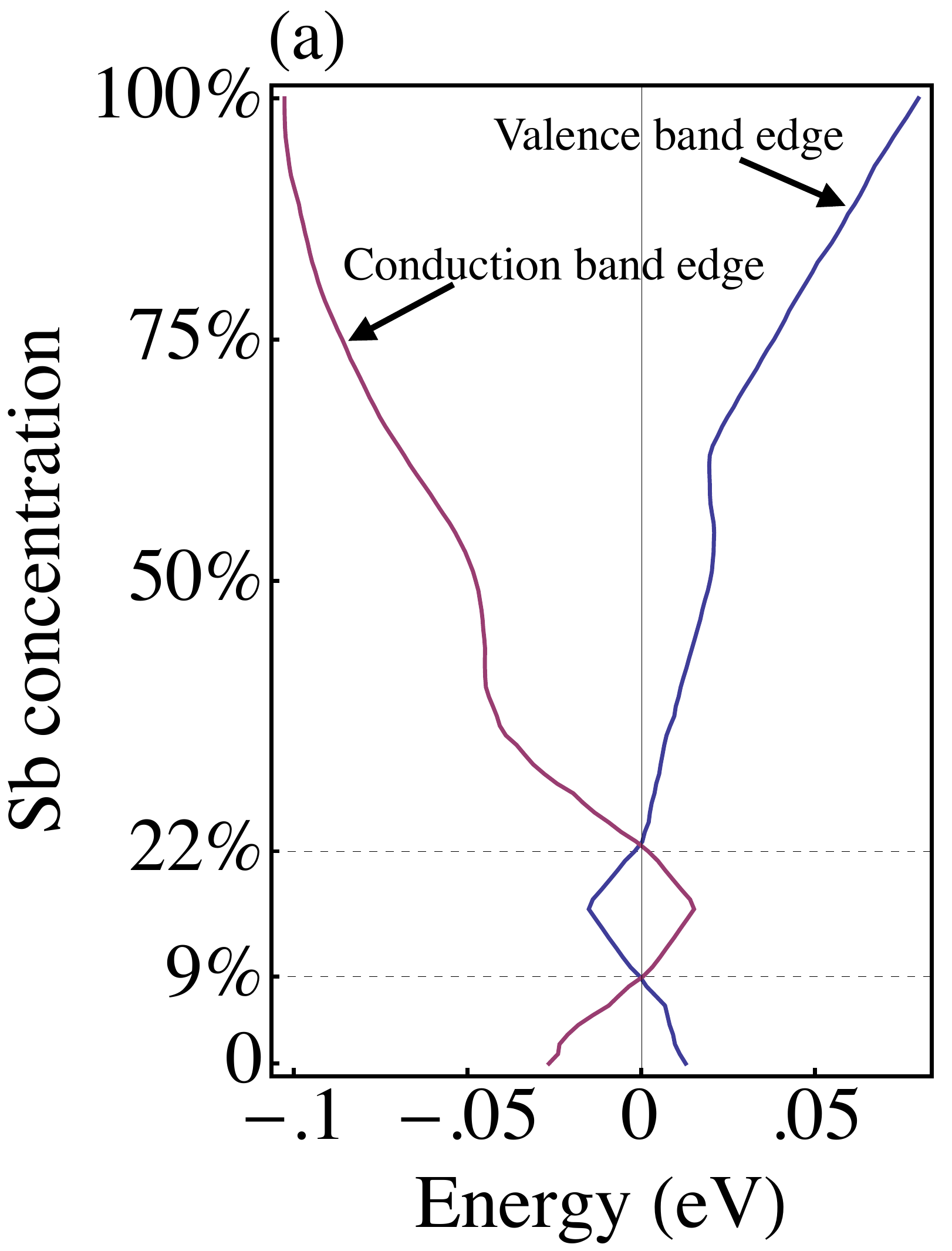} 
  \end{minipage} 
  \begin{minipage}[b]{0.49\linewidth}
    \includegraphics[width=.99\linewidth]{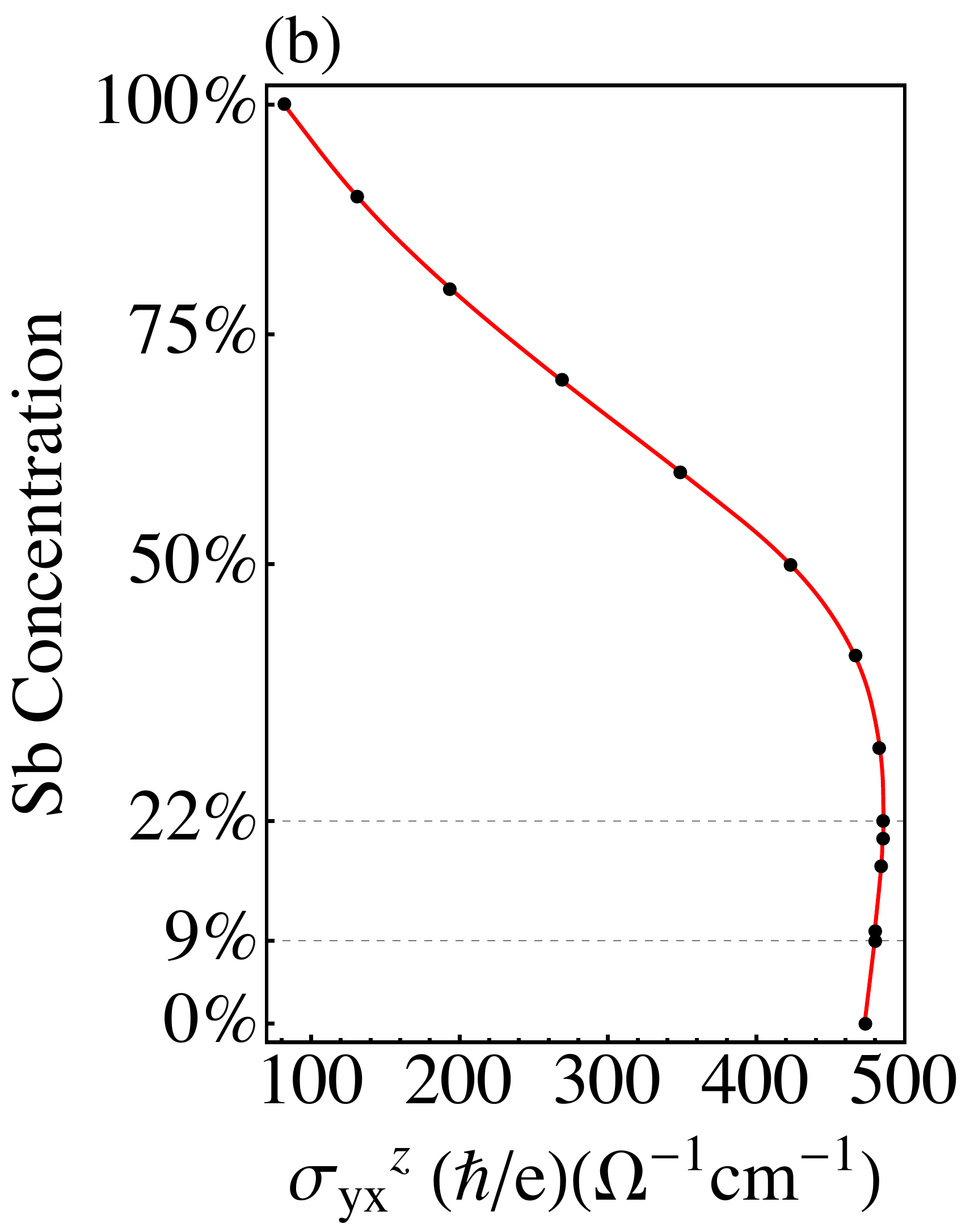} 
  \end{minipage} 
  \caption{(a) Valence band edge and conduction band edge of Bi$_{1-x}$Sb$_{x}$ as a function of antimony concentration $x$.  (b) Intrinsic spin Hall conductivity of Bi$_{1-x}$Sb$_{x}$ as a function of antimony concentration $x$. The largest spin Hall conductivity occurs near the  semimetal-semiconductor transition at 22\% antimony.}
   \label{fig:sbdependence}
\end{figure}

The electronic band structure of Bi$_{1-x}$Sb$_x$ around the Fermi energy is shown in Fig.~\ref{fig:bands} for four different compositions. The variation of the conduction and valence band edges with antimony concentration is shown in Fig.~\ref{fig:sbdependence}(a). At around 9\% antimony the band overlap disappears and a semimetal-semiconductor (SMSC) transition occurs. As the antimony concentration is increased, the valence bands shift faster than the conduction bands and an indirect gap opens, reaching a maximum of 28 meV  for Bi$_{0.83}$Sb$_{0.17}$. Up to 22\% Sb the alloy is still a semiconductor with a decreasing  band gap. At 22\% of antimony another SMSC transition occurs (Ref.~\cite{Cho1999} and references therein).

The  spin Hall conductivity in the clean static limit, evaluated as a linear response of the spin current to an electric field using a Kubo approach,  consists of a sum of the Berry curvature\cite{Guo2008}:
\al{\label{eq:shc}
\sigma_{yx}^z=\frac{e\hbar}{V}\sum_{\v k}\sum_n f_{\v k n}\Omega_n^z (\v k),}
where $e$ is the electric charge, $\hbar$ is Planck's constant, $V$ is the volume of the system, and the Berry curvature $\Omega_n^z (\v k)$ is
\al{\label{eq:curvature}
\Omega_n^z (\v k) =2\sum_{n \neq n'} {\rm Im} \frac{\big<{u_{n \v k}}|{\hat j_y^z}| {u_{n' \v k}}\big>\big< {u_{n' \v k}}|{\hat v_x}|{u_{n \v k}}\big>}{(E_{n\v k} -E_{n' \v k})^2}.
}
Here $f_{\v k n}$, the Fermi-Dirac distribution function, ensures that the sum is over all the filled bands. The electronic states $u_{n\v k}$ and $u_{n'\v k}$ are calculated from the tight-binding Hamiltonian $\hat H$ (Ref. \onlinecite{Liu1995}). The spin current and velocity operators, $\hat j_i^j$ and $\hat v_i$, are:
\al{ \label{eq:spincurrent}
\hat j_i^j=\frac{\hbar}{4}(\hat v_i\sigma_j+\sigma_j\hat v_i), 
}
\al{
\hbar \hat v_i=\nabla_{k_i}\hat H
.}

Our calculations of the spin Hall conductivity as a function of antimony concentration, Fig.~\ref{fig:sbdependence}(b), predict that both bismuth and antimony have a giant spin Hall conductivity. At room temperature bismuth has a spin Hall conductivity of 474($\hbar /e$)$(\Omega\text{cm})^{-1}$ whereas antimony's is 96($\hbar /e$)$(\Omega\text{cm})^{-1}$.  As  antimony is added to bismuth, initially the spin Hall conductivity increases, however soon it begins to drop  following the decreasing effective spin-orbit interaction in the system. There are ``hot spots'' for Berry curvature at the L and T symmetry points of the Brillouin zone for bismuth; at each of these points the curvatures are large and negative at the conduction band edge, whereas they are large and positive at the valence band edge.  As antimony is introduced to pure bismuth the conduction band edge moves away from the Fermi level, reducing the importance of the negative curvature contributions from the L point. For a small concentration of antimony this effect dominates, however at larger concentrations the band structure  changes more substantially and the Berry curvature itself decreases as the  antimony concentration is increased.

\begin{figure}[b]
  \begin{minipage}[b]{0.49\linewidth}
    \includegraphics[width=.99\linewidth]{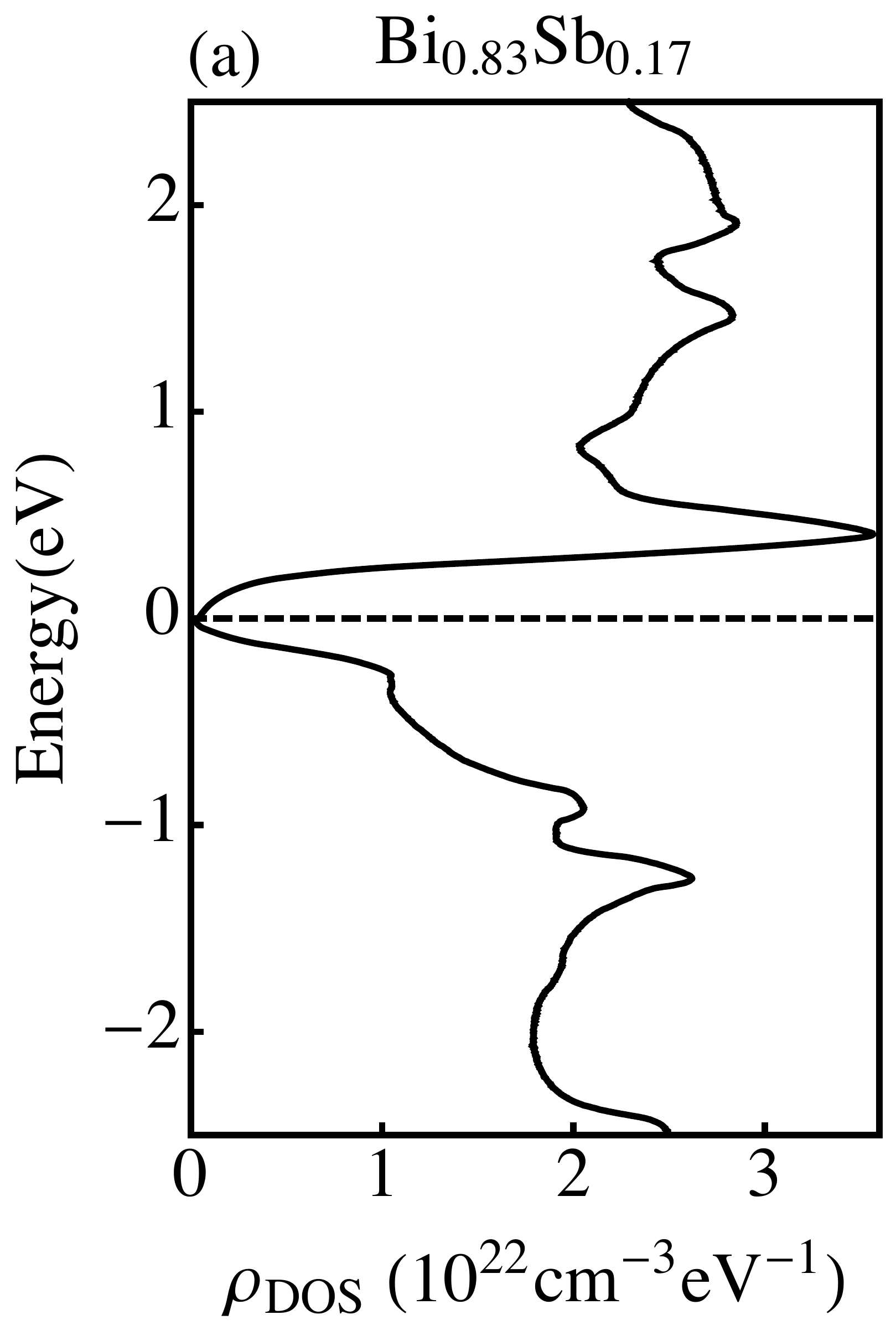} 
  \end{minipage} 
  \begin{minipage}[b]{0.49\linewidth}
    \includegraphics[width=.99\linewidth]{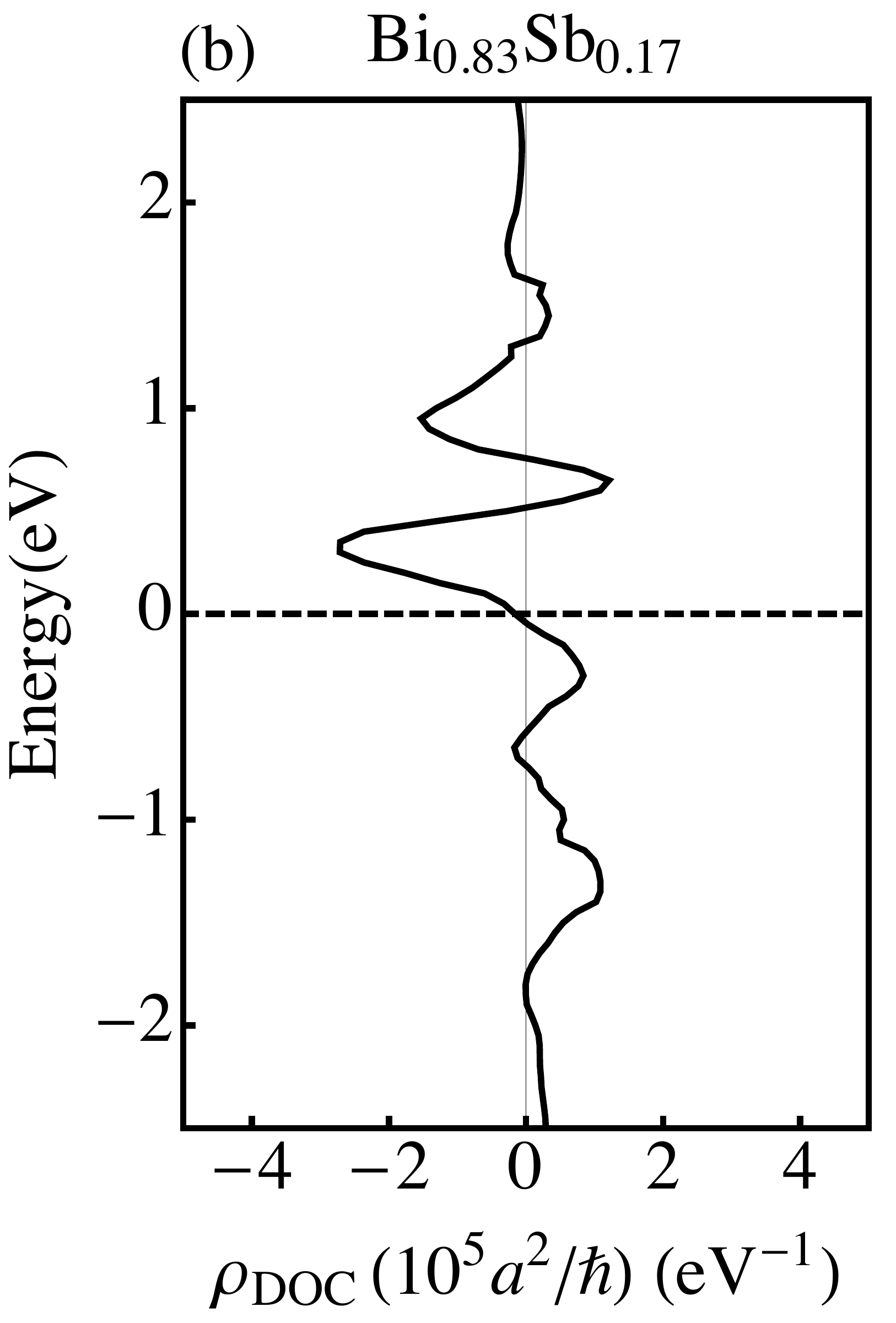} 
  \end{minipage} 
  \caption{(a) Density of states normalized per unit volume and unit energy and (b) density of curvature in the units of $(a^2/{\hbar})$ eV$^{-1}$ for Bi$_{0.83}$Sb$_{0.17}$ around Fermi energy, where $a$ is the lattice constant. The Fermi level is at 0~eV  and is indicated by the black dashed line.}
   \label{fig:densities}
\end{figure}

These features can  clearly be seen by comparing the energy dependence of the density of states with the energy dependence of the Berry curvature originating from the electronic structure.  
We have plotted the density of states around the Fermi level for Bi$_{0.83}$Sb$_{0.17}$ (the topological insulator composition with largest band gap) in Fig.~\ref{fig:densities}(a). In Fig.~\ref{fig:densities}(b) we show the density of curvature ($\rho_{\rm DOC}$), corresponding to the amount of Berry curvature per unit energy.  This quantity is useful in understanding the origin of the spin Hall conductivity and its temperature or voltage dependence, as the spin Hall conductivity can be expressed in terms of $\rho_{\rm DOC}$ as 
\al{
\sigma_{yx}^z=\frac{e\hbar}{V} \int d\epsilon \rho_{\rm DOC}(\epsilon)f(\epsilon).
}

Most of the contributions to the $\rho_{\rm DOC}$ come from the energetic regions between -2.5~eV and 2.5 eV that are shown in Fig.~\ref{fig:densities}. The valence bands at lower energy either do not possess large spin-orbit interaction or their contributions cancel; it is the presence of large spin-orbit interactions in bands close to the Fermi energy, especially those which lie on different sides of the Fermi energy, which produces the topological insulator state.

The change in sign in $\rho_{\rm DOC}$ near the Fermi energy is an additional remarkable feature that originates from the nature of the topological insulator state. The formation of a topological insulator state corresponds to the opening of a gap between strongly spin-orbit correlated states. The composition of the states at the conduction edge and the valence edge are very similar, but with opposite-sign matrix elements in Eq.~(\ref{eq:curvature}). As the Fermi energy is brought closer to the conduction edge or the valence edge, that contribution begins to dominate due to the energy denominator in Eq.~(\ref{eq:curvature}). Thus this behavior of $\rho_{\rm DOC}$, changing sign across the Fermi energy, appears to be a generic feature of topological insulators.
 
\begin{figure}
    \includegraphics[width=.99\linewidth]{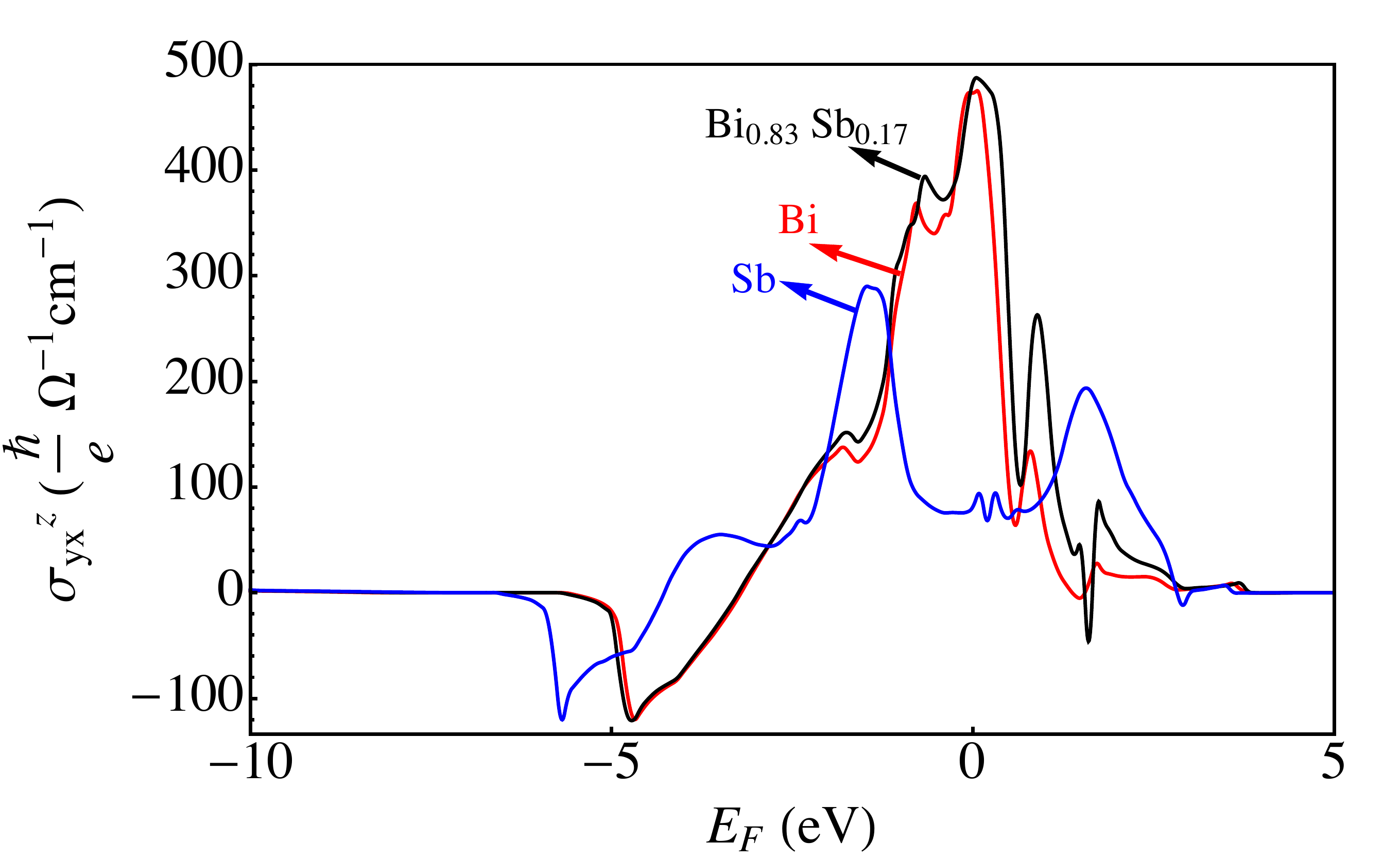} 
  \caption{The intrinsic spin Hall conductivity as a function of Fermi level for bismuth, antimony, and the topological insulator composition with the largest band gap (Bi$_{0.83}$Sb$_{0.17}$). The Fermi level for the undoped system is at 0~eV.}
   \label{fig:fermidependence}
\end{figure}



\begin{figure*}[ht!]
\begin{tabular}{ccc}
\includegraphics[width=0.66\columnwidth]{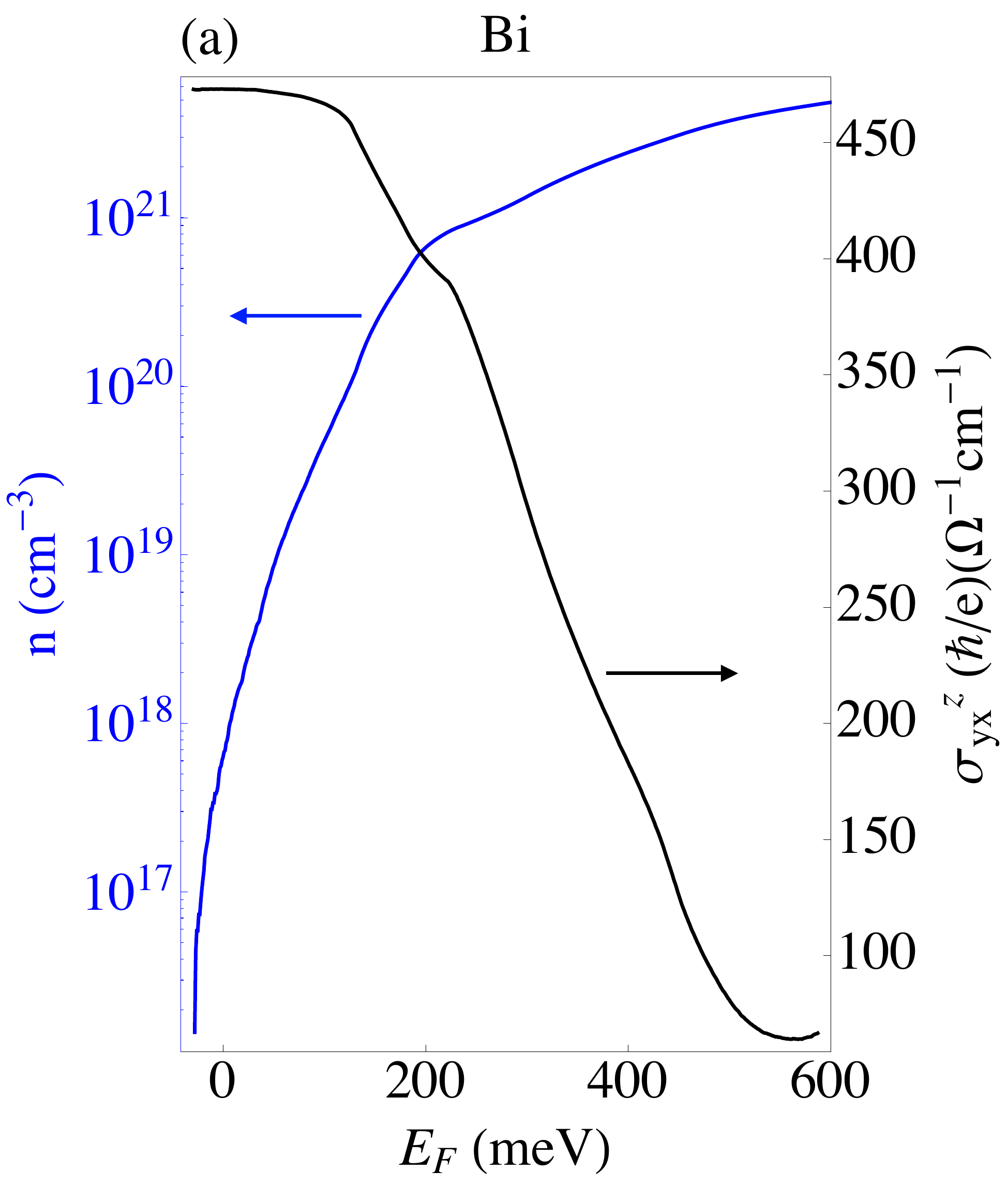}&
\includegraphics[width=0.66\columnwidth]{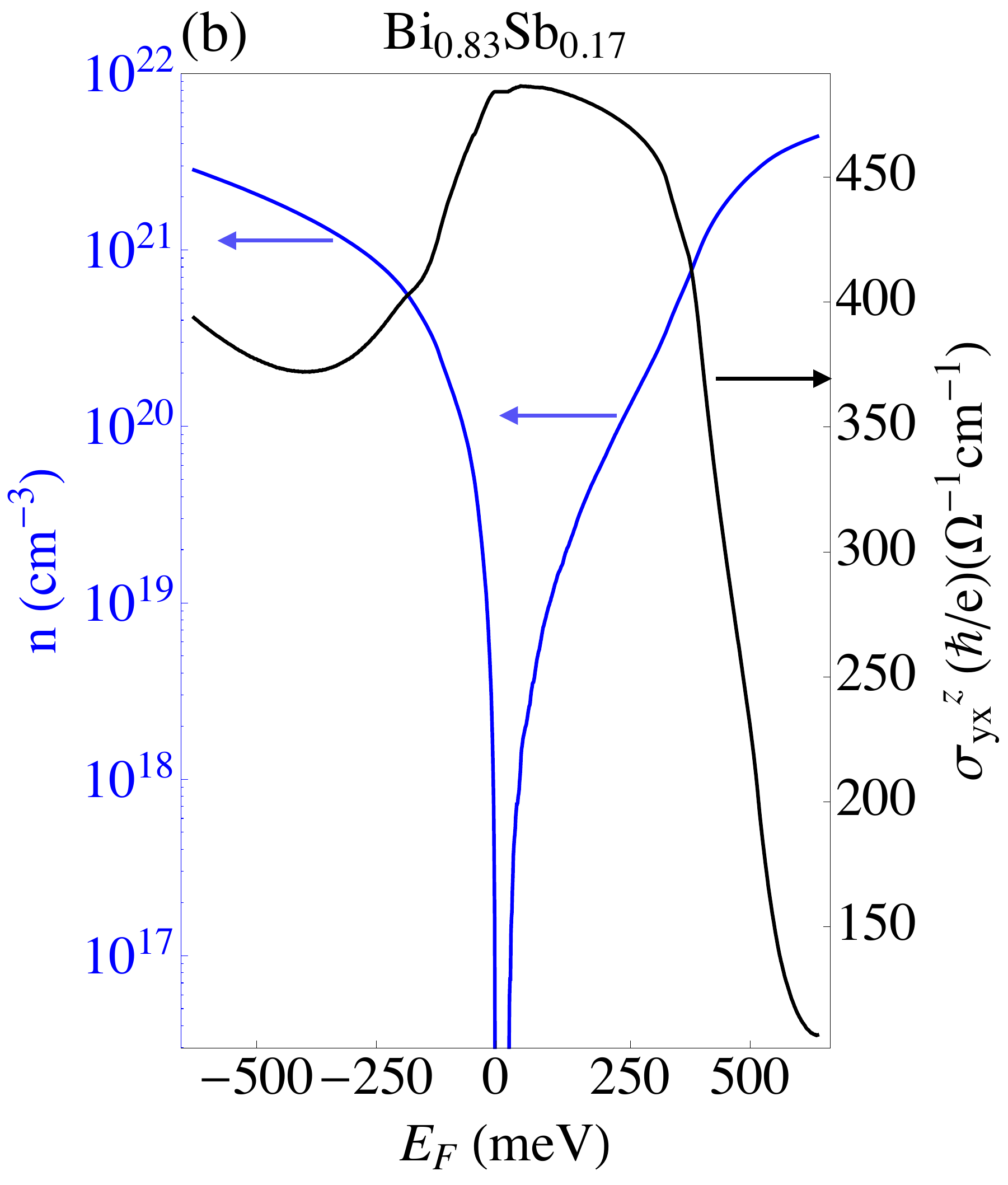}&
\includegraphics[width=0.66\columnwidth]{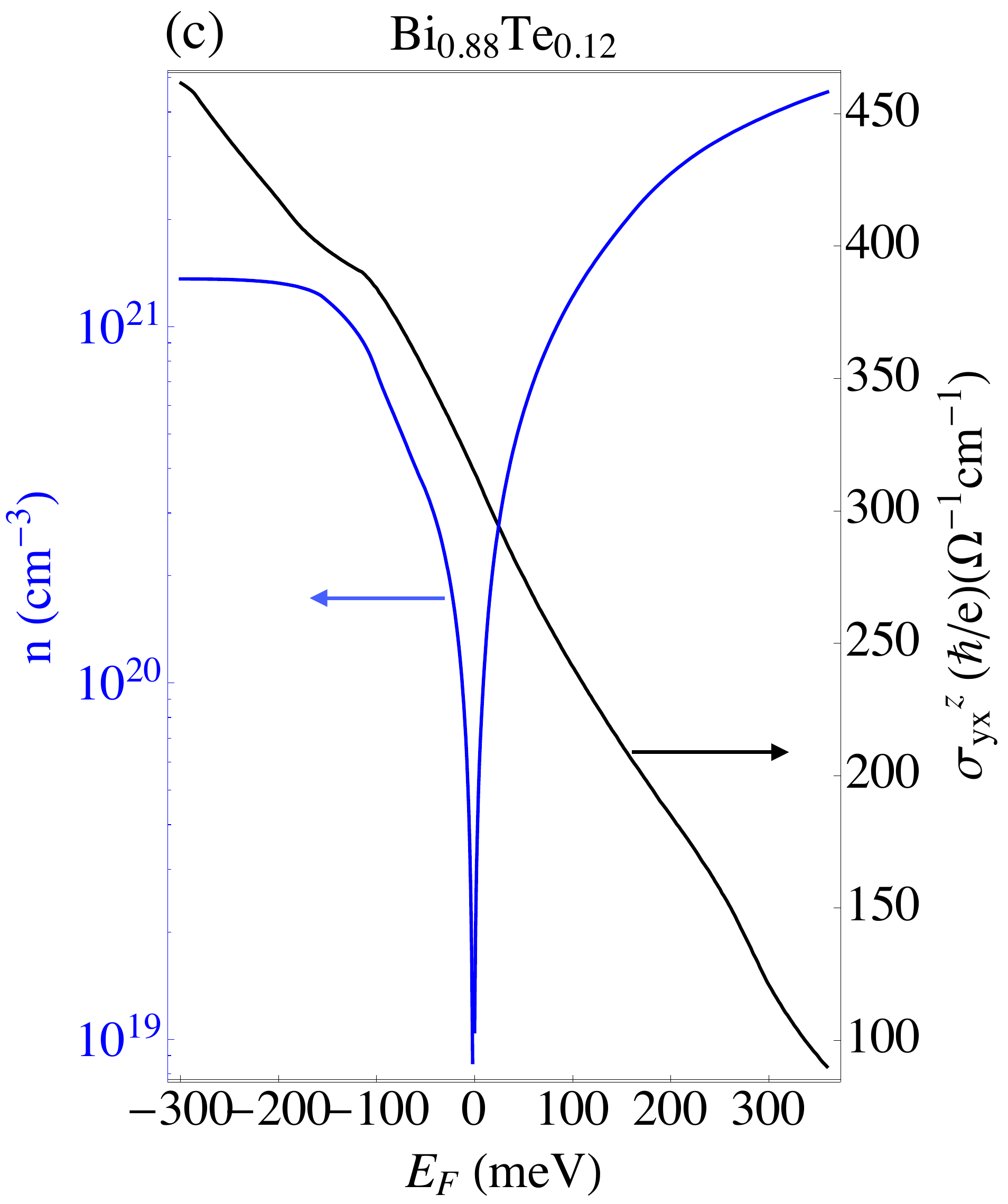}
\end{tabular}
  \caption{Gate-induced carrier densities and corresponding intrinsic spin Hall conductivities as a function of Fermi level for (a) bismuth, (b) Bi$_{0.83}$Sb$_{0.17}$ and (c) Bi$_{0.88}$Te$_{0.12}$. A range of spin Hall conductivities varying by a factor of five is achievable by doping, either via a gate or through the introduction of dopants such as Te.}
    \label{fig:carrierdensity}
\end{figure*}



We now consider the effects on the spin Hall conductivity that would come from varying the carrier concentration and Fermi energy by doping. As expected from Fig.~\ref{fig:densities}, we find a sensitive dependence of the spin Hall conductivity on the Fermi energy for both  bismuth and antimony, shown in  Fig.~\ref{fig:fermidependence}. For each material there is an optimum range for the Fermi energy which produces the largest intrinsic spin Hall conductivity. For bismuth this range  is approximately from -20 meV to +40 meV. For antimony, however, there exist several Fermi energy ranges for which the spin Hall conductivity exceeds  the intrinsic spin Hall conductivity of antimony at a Fermi energy of 0 eV. For example, a Fermi energy of -1.5 eV produces a spin Hall conductivity four times that of undoped antimony, and more than half that of bismuth (288  $\hbar$/e)($\Omega^{-1}$cm$^{-1}$). A Fermi energy of 1.5 eV produces a spin Hall conductivity somewhat less, but still more than twice that of undoped antimony $\sigma_{yx}^z= 188$ ($\hbar$/e)($\Omega^{-1}$cm$^{-1}$). By comparison the topological insulator material Bi$_{0.83}$Sb$_{0.17}$ does not possess a larger spin Hall conductivity than bismuth, and in fact its spin Hall conductivity as a function of Fermi energy is very similar to that of bismuth. We thus note that the dominant contribution to the spin Hall conductivity comes from the large spin-orbit interaction in the materials, rather than the topological character of the band structures.  For sufficiently low Fermi energy ($< -6$~eV) the spin Hall conductivity vanishes because all the bands are entirely full or entirely empty. The presence of this band gap deep within the valence structure of bismuth or antimony is a property of the electronic structure model Hamiltonian for these systems (Ref.~\onlinecite{Liu1995}).

From the Fermi-energy dependence of the spin Hall conductivity and the density of states of the materials  we predict the change in the spin Hall conductivity with carrier density, shown in Fig.~\ref{fig:carrierdensity}. We expect this change would be achieved through accumulation or depletion via an electrical gate in a field-effect transistor device. The change in carrier density is plotted as a function of the change in carrier density (electron or hole). The equilibrium carrier density of  semimetallic bismuth is 3.1x10$^{17}$cm$^{-3}$, which is many orders of magnitude lower than the carrier concentration of typical metals.  For Bi$_{\rm 0.83}$Sb$_{\rm 0.17}$ the equilibrium bulk carrier concentration vanishes at low temperature.  As the Fermi level is changed by a gate voltage the materials exhibit more metallic behavior.   Changes in carrier concentration modify the spin Hall conductivity by approximately a factor of five, suggesting that gate-tuning the spin Hall conductivity of such materials is possible. For bismuth there is little change in the spin Hall conductivity for an initial change of the Fermi energy by 150 meV. Instead of gate-tuning to this point it should be possible to dope the material with a group-VI dopant such as Te. For a Te concentration of 12\% the spin Hall conductivity lies in between the upper and lower extremes, producing the largest tuning range with voltage. Thus we present in Fig.~\ref{fig:carrierdensity}(c) the carrier-dependence of the spin Hall conductivity for Bi$_{\rm 0.88}$Te$_{\rm 0.12}$. We note that this doping consists of adding Te to the crystal structure of Bi, not shifting to the crystal structure of Bi$_2$Te$_3$.  As the longitudinal conductivity of these materials will change as well with a change in the Fermi energy we expect that the spin Hall angle, defined as the ratio of the spin Hall conductivity to the longitudinal conductivity, could be substantially varied as well.

We have calculated the intrinsic spin Hall conductivity for bismuth, antimony, and BiSb alloys, using a Berry's curvature technique. The electronic structures are described by a three-nearest-neighbor tight-binding Hamiltonian, within which the alloys are treated in a virtual crystal approximation.  
We find little difference in the magnitude of the spin Hall conductivity between bismuth and the topological insulator material Bi$_{0.83}$Sb$_{0.17}$. However the longitudinal conductivity will vary considerably between these two materials, so that the spin Hall angle of Bi$_{0.83}$Sb$_{0.17}$ should greatly exceed that of bismuth. Calculations of the Fermi level dependence of the spin Hall conductivity suggests that substantial (factor of five) gate tuning of the spin Hall conductivity is possible. Bismuth, antimony and \BiSb alloys with large spin-orbit couplings exhibit robust intrinsic spin Hall conductivities, larger than conventional semiconductors and metals with large spin Hall conductivity. Bismuth, antimony and bismuth-antimony alloys are thus promising candidates for transverse spin current generation and  spintronic applications.

This work was supported in part by C-SPIN, one of six centers of STARnet, a Semiconductor Research Corporation program, sponsored by MARCO and DARPA.

\bibliography{central-bibliography}
\end{document}